\documentclass[preprint,12pt]{elsarticle}
\usepackage[brazil]{babel}
\usepackage{amsmath,amsthm,amsfonts,amssymb}
\usepackage{dsfont}
\journal{Physics Letters B}

\begin{document}
\begin{frontmatter}
\title{Kinematics and dynamics in a Bipartite-Finsler spacetime}
\author{J. E. G. Silva, C. A. S. Almeida}
\address{Departamento de F\'{i}sica - Universidade Federal do Cear\'{a} \\ C.P. 6030, 60455-760
Fortaleza-Cear\' {a}-Brazil}

\begin{abstract}
We study some properties of a recently proposed local Lorentz Violating Finsler geometry, the so-called Bipartite space.
This anisotropic structure deforms the causal null surface to an elliptic cone and provides an
anisotropy to the inertia. We obtain the new modified dispersion relations and the geodesic equation for a massive particle.
For a weak directional-dependence we find the dynamical and interaction terms analogous to the gravitational sector of the Standard Model Extension.
\end{abstract}
\end{frontmatter}



\section{Introduction}
In the regime of a quantum gravity many theories expects the space-time be no longer locally isotropic. For the String Theory, the tensor fields
can spontaneously break the Lorentz symmetry by assuming a definite vacuum expected value
\cite{Kostelecky:1988zi,Kostelecky:1990pe}. An effective theory comprising this effect, proposed by Kosteleck\'{y} and collaborators, is
called the Standard Model Extension (SME) \cite{Colladay:1998fq}. On the other hand,
the existence of a minimal length breaks the Lorentz symmetry and leads to the Doubly Special Relativity (DSR) \cite{Magueijo:2001cr}. The Very Special Relativity (VSR), where the symmetry group of the space-time is the subgroup $SIM(2)$ of the Lorentz group, also violates the Lorentz symmetry by allowing a spurious vector field similar to the aether model \cite{Cohen:2006ky,Jacobson:2000xp}.

In order to study Lorentz violating gravitational effects, i.e., to extend the break of the Lorentz symmetry to curved space-times, some of these models dismissed the Riemannian background (locally isotropic) for a Finsler geometry (locally anisotropic) approach. In Finsler geometry, the length of a curve is measured using a general function of the position $x\in M$ and direction $y\in T_{x}M$, called Finsler function $F(x,y)$, in the form \cite{chern}
\begin{equation}
\label{finslerarclengh}
 s=\int_{I}{F(x,\dot{x})d\tau},
\end{equation}
where $\tau\in I$ is an affine parameter and $\dot{x}=\frac{dx}{d\tau}$ is a tangent vector. Physically it means that the
measurement of the proper time or the lengths are now directional-dependents \cite{Pfeifer:2011xi,Pfeifer:2011tk}. This is a fundamental form to include the local Lorentz violation into
the space-time itself and to the fields and particles living on it.

A particular choice of the Finsler function defines a specific new geometry. In Riemannian geometry $F(x,y)=\sqrt{g_{\mu\nu}y^{\mu}y^{\nu}}$. For $F(x,y)=\sqrt{g_{\mu\nu}y^{\mu}y^{\nu}} + a_{\mu}y^{\mu}$ we have the Randers space \cite{randers} whose vector $a_{\mu}$, besides providing the local anisotropy, can explain both the dark matter and dark energy \cite{Chang:2008yv,Chang:2009pa}. For the VSR the Finsler function is given by
$F(x,y)=(n_{\rho}y^{\rho})^{b}(g_{\mu\nu}y^{\mu}y^{\nu})^{\frac{(1-b)}{2}}$ \cite{Gibbons:2007iu}, which defines the Bogoslovsky space \cite{Bogoslovsky:1998wa,Bogoslovsky:2004rp}. The spurious vector field $n_{\rho}$ is a possible source for the dark energy and the inflation \cite{Kouretsis:2008ha,Kouretsis:2010vs}. The Finslerian structure  of the DSR yields to its modification of the dispersion relation (MDR) $P_{\mu}P^{\mu}=-(1-\lambda P_{0})^{2}m^{2}$ \cite{Girelli:2006fw}.

The Modifications of the dispersion relations are usual features of Finsler-based theories. Indeed, given a Finsler Function is possible to define a symmetric tensor called the Finsler metric $\mathbf{g}^{F}(x,y)=g_{\mu\nu}^{F}(x,y)dx^{\mu}\otimes dx^{\nu}$ by \cite{chern}
\begin{equation}
g^{F}_{\mu\nu}(x,y)=\frac{1}{2}\frac{\partial^{2}F^{2}}{\partial y^{\mu}\partial y^{\nu}}.
\end{equation}
Note that the directional-dependence is already encoded in the metric tensor. The square of the vector is defined as $||y||^{2}=g_{\mu\nu}^{F}(x,y)y^{\mu}y^{\nu}$ that allows non-quadratic terms \cite{Girelli:2006fw,Chang:2008zzb,Vacaru:2010fa,Vacaru:2010rd}.

The SME also posses a Finsler-based structure. In fact, the curved extension of SME is made by a spontaneous symmetry breaking mechanism since the explicit Lorentz violation breaks the Bianchi identities \cite{Kostelecky:2003fs,Maluf:2013nva}. A proposal to overcome this is done through some Finsler geometries \cite{Kostelecky:2003fs,Kostelecky:2010hs,Kostelecky:2011qz}. A spin-$1/2$ fermion with Lorentz violating terms has modified dispersion relations that can be associated with a point-particle moving in a Finsler space-time with a Finsler function $F(x,y)=\sqrt{|g_{\mu\nu}(x)y^{\mu}y^{\nu}|} + a_{\mu}y^{\mu} + \sqrt{s_{\mu\nu}(x)y^{\mu}y^{\nu}}$ which extends the Randers metric \cite{Kostelecky:2010hs,Kostelecky:2011qz}. For $a_{\mu}=0$ the space turns to be a new Finsler structure called the Bipartite space \cite{Kostelecky:2012ac}. The Randers term is responsible for the CPT-odd effects whereas the Bipartite term belongs to the CPT-even sector \cite{Kostelecky:2011qz,Kostelecky:2012ac}. Note that this SME-based Finsler geometry can be understood as a small perturbation over the local Lorentz invariant geometry. However, the Finslerian approach has the advantage of treat the geometry naturally anisotropic without any external field. Further, it can also provides torsion as a natural effect.

In this work we explore some basic features and find new interesting properties of the Bipartite space. Since this Finsler structure was recently proposed, there are many open points to address. The main goal here is to compare the properties of this spacetime with other Lorentz-violating models and with anisotropic media. In this regard, we propose a new perspective that the Bipartite space provides some effects analogous to a background tensor field on a Lorentzian space. In section (\ref{Kinematics}) we show that the causal surface is an elliptic cone. Another new result is that the time difference measured by inertial observers is directional-dependent.
In section (\ref{Particle dynamics}) we obtain an anisotropic momentum and we study the corresponding MDR of a free particle. Moreover, we find that a free particle in this geometry moves analogously to a particle on a Lorentzian spacetime with a background field,due to a new anisotropic term in the geodesic equation. The section (\ref{Dynamics of the geometry}) is devoted to the first step in study the dynamics of the Bipartite space. Indeed, the analysis presented by Kostelecky \textit{et al} \cite{Kostelecky:2010hs,Kostelecky:2011qz,Kostelecky:2012ac} is performed for a fixed background geometry. We argue that a dynamics for the Finslerian metric $\mathbf{g}^{F}$ can be divided into a dynamics for the Lorentzian metric $\mathbf{g}$ and for the Bipartite tensor $\mathbf{s}$. For the weak directional-dependence limit, where the dependence of the geometry on the direction is taken only on the components of the tensors and for tiny values of the Bipartite tensor, we show that a Finslerian Einstein-Hilbert (EH) action can be split out into a Lorentzian EH action plus some coupling terms between the Lorentizian metric and the Bipartite tensor similar to those of the SME.


\section{Kinematics}
\label{Kinematics}
Consider a space-time $M$ endowed with a Lorentzian metric
$\mathbf{g}$\footnote{We adopt the mostly plus convention $(-,+,+,+)$ for the metric signature.} $\in T^{*}M$ and a symmetric tensor $\mathbf{s}\in T^{*}M$ whose Finsler function is given by \cite{Kostelecky:2011qz,Kostelecky:2012ac}
\begin{eqnarray}
\label{Bipartitefinslerfunction}
F(x,y)_{B} & = & \alpha(x,y) + \sigma(x,y)\nonumber\\
       & = & \sqrt{|g_{\mu\nu}(x)y^{\mu}y^{\nu}|} + \xi\sqrt{s_{\mu\nu}(x)y^{\mu}y^{\nu}},
\end{eqnarray}
$0\leq \xi \leq 1$ is a constant controlling the local Lorentz violation \footnote{The constant $\xi$, not present in the Kostelecky works \cite{Kostelecky:2011qz,Kostelecky:2012ac}, allow us to ensure that $0< \sigma < 1$.}. The tiple $\mathcal{B}=(M,\mathbf{g},\mathbf{s})$ is called a Bipartite space. Hereupon we shall consider the Bipartite tensor $\mathbf{s}$ with mass dimension $[\mathbf{s}]=2$ in four dimensions what leads $\xi$ to have mass dimension $[\xi]=-1$.
Note that unlike the Randers function, the Bipartite Finsler function is parity invariant, i.e., $F_{B}(x,-y)=F_{B}(x,y)$.

An interesting choice for the Bipartite tensor is \cite{Kostelecky:2011qz,Kostelecky:2012ac}
\begin{equation}
\label{Bipartitetensor}
 \mathbf{s}_{b}=\mathbf{b}\otimes\mathbf{b}-b^{2}\mathbf{g},
\end{equation}
where $b^{2}=\mathbf{g}(\mathbf{b},\mathbf{b})$. This space-time is called a $b$-space \cite{Kostelecky:2011qz}. It is worthwhile to say that the Bipartite tensor in (\ref{Bipartitetensor}) is analogous to the Lorentz-violating tensor field in the bumblebee model \cite{Kostelecky:2003fs,Maluf:2013nva}.

Following \cite{Kostelecky:2012ac} we define an idempotent transformation $\mathfrak{s}:TM\rightarrow TM$ by $\mathfrak{s}=s^{\mu}_{\nu}e_{\mu}\otimes dx^{\nu}$, having a non-zero eigenvalue $\zeta$ such that $\mathfrak{s}^{2}=\zeta \mathfrak{s}$. The eigenvalue $\zeta$ has mass dimension $[\zeta]=M^{2}$ and for the $b$-space  $\zeta=b^{2}$.

In the following sections we obtain new and intriguing features of the Biparite spacetime.

\subsection{Causal structure}
The first new properties is that the Bipartite structure deforms the usual Lorentz light cone. Indeed, since $ds=F(x,y)dt$, the null interval satisfies
$F(x,\dot{x})=0$ what yields to
\begin{equation}
(\eta_{\mu\nu}-\xi^{2}s_{\mu\nu})\dot{x}^{\mu}\dot{x}^{\nu}=0.
\end{equation}
Therefore, for \textbf{$x=(X_{0},X_{1},X_{2},X_{3})\in M$} the causal surface is the cone
\begin{equation}
\label{generalcausalsurface}
-(1+\xi^{2}s_{00})X_{0}^{2} + (1-\xi^{2}s_{ij})X^{i}X^{j} - 2\xi^{2}s_{0i}X^{0}X^{i}=0.
\end{equation}

Consider a base $(\mathbf{\dot{x}},\mathbf{e}_{1},\mathbf{e}_{2},\mathbf{e}_{3})\in T^{*}M$ formed with the mutual eigenvectors of $\mathbf{s}$ and $\mathbf{\eta}$. The Bipartite tensor is written as $\mathbf{s}=\lambda_{0}\mathbf{\dot{x}}\otimes \mathbf{\dot{x}} + \sum_{i=1}^{3}{\lambda_{i}\mathbf{e}_{i}\otimes\mathbf{e}_{i}}$.
The matrix of $\mathbf{s}$ in this base is given by $s=diag(\lambda_{0},\lambda_{1},\lambda_{2},\lambda_{3})$, where $(\lambda)_{i}$ are the set of eigenvalues of $\mathbf{s}$.
Thus, we find that the causal surface turns to be an elliptic cone of form
\begin{equation}
(1-\xi^{2}\lambda_{1})x_{1}^{2}+(1-\xi^{2}\lambda_{2})x_{2}^{2}+(1-\xi^{2}\lambda_{3})x_{3}^{2}-(1+\xi^{2}\lambda_{0})x_{0}^{2}=0.
\end{equation}
As interesting new consequence of this space-time is that since the generatrices have different slopes, the light moves with different speeds depending on the direction. In order to avoid causal issues, as
superluminal velocities, we impose the condition
\begin{eqnarray}
\label{causalcondition}
\frac{\partial x_{i}}{\partial x_{0}} = \sqrt{\frac{1-\xi^{2}\lambda_{i}}{1+\xi^{2}\lambda_{0}}}\leq 1 & \Rightarrow & \lambda_{0}+\lambda_{i}\geq 0.
\end{eqnarray}
Further, from $s_{\mu\nu}\dot{x}^{\mu}\dot{x}^{\nu}\geq 0 \Rightarrow \lambda_{0}\geq 0$. These conditions on the Bipartite tensor resembles the weak energy condition for the stress-energy tensor in General Relativity \cite{wald}.

Another important new causal property is that the Bipartite tensor $\mathbf{s}$ also changes the time measured by inertial observers. Indeed, consider a massive particle with 4-velocity $\mathbf{\dot{x}}$. For the Minkowsky metric $\mathbf{g}=\mathbf{\eta}$, in the rest frame the 4-velocity is $\dot{x}=\frac{d\mathbf{x}}{d\tau}=(1,\vec{0})$, where $\tau$ is the proper time. The interval takes the form $ds^{F}=(1+\xi\sqrt{s_{00}})d\tau$. In another inertial frame, moving with velocity $\vec{v}$ in respect to the first, the 4-velocity is given by $\mathbf{\dot{x}}'=\frac{d\mathbf{x}}{dt}=(1,\vec{v})$ what yields to the interval $ds'=\left(\sqrt{1-v^{2}} + \xi\sqrt{s_{\mu\nu}\dot{x}^{'\mu}\dot{x}^{'\nu}}\right)dt$. From $ds=ds'$ the relation between $d\tau$ and $d\tau=\frac{dt}{\gamma(\vec{v},\mathbf{s})^{F}}$ is given by
\begin{equation}
\label{bipartitegamma}
\gamma(\vec{v},\mathbf{s})^{F} = \frac{1+\xi\sqrt{s_{00}}}{\sqrt{1-v^{2}}+\xi\sigma(x,\dot{x'})}.
\end{equation}
For the $b$-space $\gamma(\vec{v},\mathbf{s})^{F} = \frac{1+\xi \sqrt{||\vec{b}||}}{1-v^{2}+\xi\sqrt{||\vec{b}||^{2}+(\vec{b}\cdot\vec{v})^{2}}}$. Thus, the time difference depends on the relative direction in respect to the background vector $\vec{b}$. This result suggests an analogy between the Bipartite structure and the vector Lorentz-violating models, as the aether \cite{Jacobson:2000xp} or the bumblebee model \cite{Kostelecky:2003fs}.


\section{Particle dynamics}
\label{Particle dynamics}
Now let us study the dynamics of a free particle moving on a Bipartite space. The action functional for a point particle of mass $m$ is given by \cite{Pfeifer:2011tk,Chang:2008yv,Girelli:2006fw}
\begin{equation}
\label{Bipartiteaction}
S^{F}=m\int_{I}{F(x,\dot{x})d\tau},
\end{equation}
Therefore, the Lagrangian for the point particle is
\begin{eqnarray}
\label{BipartiteLagrangian}
L^{F} & = & mF(x,\dot{x})\nonumber\\
  & = & m\left(\sqrt{g_{\mu\nu}(x)\dot{x}^{\mu}\dot{x}^{\nu}} + \xi\sqrt{s_{\mu\nu}(x)\dot{x}^{\mu}\dot{x}^{\nu}}\right)
\end{eqnarray}
whose canonical conjugate Finslerian 4-momentum is given by
\begin{eqnarray}
\label{finslermomentum}
P_{\mu}^{F} &   =   & \frac{\partial L^{F}}{\partial \dot{x}^{\mu}}\nonumber\\
             &   =   &   M_{\mu\nu}(x)\dot{x}^{\nu},
\end{eqnarray}
where $M_{\mu\nu}(x)=M_{\nu\mu}(x)$ is given by
\begin{equation}
\label{inertiatensor}
M_{\mu\nu}(x)=m\left(g_{\mu\nu}(x)+\frac{\xi}{\sigma}s_{\mu\nu}(x)\right)
\end{equation}
and it can be understood as an inertia tensor. A canonical momentum similar to found in eq.(\ref{finslermomentum}) was obtained for the $b$-space \cite{Kostelecky:2010hs}. The anisotropy of the inertia is a intriguing result
also present in the Bogoslosky model \cite{Bogoslovsky:1998wa,Bogoslovsky:2004rp,Kouretsis:2010vs}. However, here the anisotropic inertia arises from a symmetric and geometric tensor.

The equation (\ref{finslermomentum}) can be rewritten as the sum $P_{\mu}^{F}=P_{\mu}+\xi\tilde{P}_{\mu}$, where $P_{\mu}=mg_{\mu\nu}\dot{x}^{\nu}$ is
the usual Lorentzian 4-momentum and $\tilde{P}_{\mu}=\frac{m}{\sigma}s_{\mu\nu}\dot{x}^{\nu}$ is a 4-momentum arising due to the anisotropy of the space-time. An anisotropic momentum of form $P^{F}_{\mu}=m(g_{\mu\nu}\dot{x}^{\nu}+\xi a_{\mu}\dot{x}^{\nu})$ has also been obtained by Randers \cite{randers}, which is analogous to a 4-momentum of a particle moving on a Lorentzian space-time minimally coupled with a background electromagnetic vector potential $\mathbf{a}$. Therefore, we interpret the momentum (\ref{finslermomentum}) as a coupling between a point particle in a
Lorentzian space-time with a background tensor field $\mathbf{s}$.


\subsection{Modified dispersion relations}
The Finsler metric for the
Bipartite space is given by \cite{Kostelecky:2012ac}

\begin{equation}
\label{Bipartitefinslermetric}
 g_{\mu\nu}^{F}(x,y) = \frac{F}{\alpha}g_{\mu\nu} + \xi\left(\frac{F}{\sigma}s_{\mu\nu}-\alpha\sigma k_{\mu}k_{\nu}\right),
\end{equation}
where, $k_{\mu}=\frac{1}{\alpha}\frac{\partial\alpha}{\partial y^{\mu}}-\frac{1}{\sigma}\frac{\partial\sigma}{\partial y^{\mu}}$. We define the unit vector
$\tilde{l}_{\mu}=\frac{\partial \alpha}{\partial y^{\mu}}=\frac{1}{\alpha}g_{\mu\beta}y^{\beta}$ \cite{chern} and the vector $\hat{l}_{\mu}=\frac{\partial \sigma}{\partial y^{\mu}}=\frac{1}{\sigma}s_{\mu\beta}y^{\beta}$.

The inverse Bipartite metric is given by \cite{Kostelecky:2012ac}
\begin{equation}
\label{inverseBipartitefinslermetric}
 g^{F\mu\nu}(x,y) = \frac{\alpha}{F}g^{\mu\nu} - \frac{\xi\alpha^{2}}{FS}\Big[s^{\mu\nu} - \left(\frac{S}{F}\right)^{2}\tilde{l}^{\mu}\tilde{l}^{\nu} + \frac{S}{F}\tilde{l}^{(\mu}\hat{l}^{\nu)} - \hat{l}^{\mu}\hat{l}^{\nu}\Big],
\end{equation}
where, $S=\sigma + \xi\zeta\alpha$. Thus, using the inverse Finslerian metric (\ref{inverseBipartitefinslermetric}) to measure the length of vector, the square of the Finslerian 4-momentum is
\begin{eqnarray}
\label{squaredmomentum}
||P^{F}||^{2}    & = & g^{F\mu\nu}(x,P^{F})P^{F}_{\mu}P^{F}_{\nu}\nonumber\\
                 & = & \Big[\frac{\alpha}{F}g^{\mu\nu} - \frac{\xi\alpha^{2}}{FS}s^{\mu\nu}\Big]P_{\mu}^{F}P_{\nu}^{F}\nonumber\\
        & + & \frac{\xi\alpha^{2}}{FS}\Big[\left(\frac{S}{F}\right)^{2}\frac{g^{\mu\beta}g^{\nu\epsilon}}{\alpha^{2}} - \frac{S}{F}\frac{g^{\mu\beta}s^{\nu\epsilon}}{\alpha\sigma} + \frac{s^{\mu\beta}s^{\nu\epsilon}}{\sigma^{2}}\Big]P_{\mu}^{F}P_{\nu}^{F}P_{\beta}^{F}P_{\epsilon}^{F}.
\end{eqnarray}
The first line of equation (\ref{squaredmomentum}) has quadratic terms in the momentum whereas the second line provides
quartic terms, similar to other Finsler space-times \cite{Chang:2008yv,Bogoslovsky:1998wa,Girelli:2006fw,Chang:2008zzb,Vacaru:2010fa,Vacaru:2010rd,Kostelecky:2011qz}.
For the Minkowsky space-time $\mathbf{g}=\mathbf{\eta}$, the Bipartite 4-momentum satisfying $||P^{F}||^{2}=-m^{2}$ yields to
\begin{eqnarray}
\label{modifieddispersionsrelations}
\frac{\alpha}{F}P^{F\mu}P_{\mu}^{F} - \xi\frac{P^{F\mu}P_{\mu}^{F}}{FS}\Big[\left(\frac{S}{F}\right)^{2}P^{F\mu}P_{\mu}^{F} + \frac{S\alpha\sigma}{F}\Big]=-m^{2}
\end{eqnarray}
where $\alpha=\alpha(x,P^{F})=\sqrt{\eta^{\mu\nu}P_{\mu}^{F}P_{\nu}^{F}}=\sqrt{|-E^{2}+P^{2} + 2\xi P^{\mu}\tilde{P}_{\mu} + \xi^{2}\tilde{P}^{\mu}\tilde{P}_{\mu}|}$ and $\sigma=\sigma(x,P^{F})=\sqrt{s^{\mu\nu}P_{\mu}^{F}P_{\nu}^{F}}$.
A similar result is present in the
Randers space \cite{Chang:2008zzb}. Expanding the eq.(\ref{modifieddispersionsrelations}) until first order in $\xi$ we obtain
\begin{equation}
E^{2}-P^{2}+\xi(\alpha\sigma - 2P^{\mu}\tilde{P}_{\mu}) = m^{2}
\end{equation}
At the rest frame, $(1 + \xi s_{00})E^{2} -2\xi\tilde{P}_{0}E - m^{2}= 0$ whose solution is
\begin{equation}
E=m(1+\xi\sqrt{s_{00}})^{-\frac{1}{2}} + \xi \frac{\tilde{P}_{0}}{(1+\xi\sqrt{s_{00}})}.
\end{equation}
Thus, for the $b$ space, where $s_{00}=||\vec{b}||^{2}$, a time-like vector $b=(b_{0},\vec{0})$ describing an aether model does not change the relation energy-mass.

After the identification $P_{\mu}=-i\partial_{\mu}$, a scalar field will satisfy the equation
\begin{equation}
\{\Box + \xi[\hat{\alpha}\hat{\sigma} - 2i\tilde{P}^{\mu}\partial_{\mu}]\}\Phi=m^{2}\Phi,
\end{equation}
where
\begin{eqnarray}
\tilde{\alpha}  &   =   &   \sqrt{\Box - 2i\xi \tilde{P}^{\mu}\partial_{\mu} + \xi^{2}\tilde{P}^{\mu}\tilde{P}_{\mu}\mathds{1}}\nonumber\\
\hat{\sigma}    &   =   &   \sqrt{s^{\mu\nu}\partial_{\mu}\partial_{\nu} - 2i\xi s^{\mu\nu}\tilde{P}^{\mu}\partial_{\mu} + \xi^{2}s^{\mu\nu}\tilde{P}_{\mu}\tilde{P}_{\nu}\mathds{1}}.
\end{eqnarray}
which is similar to the SME Lorentz Violating equation \cite{Kostelecky:1988zi,Kostelecky:1990pe,Kostelecky:2003fs}.


\subsection{Geodesic motion}
Now let us analyze the equation of motion of a particle. In order to the world-line be an extremum the action (\ref{Bipartiteaction}) it must satisfies the geodesic equation \cite{chern}
\begin{equation}
\label{Bipartitesecondlaw}
M^{\mu}_{\nu}\ddot{x}^{\nu} = F^{\mu}.
\end{equation}
where, $F^{\mu}=-m[\gamma^{\mu}_{\nu\beta} + \xi(\tilde{\gamma}^{\mu}_{\nu\beta} + \partial_{\nu}(\sigma^{-1}))\delta^{\mu}_{\beta}]\dot{x}^{\nu}\dot{x}^{\beta}$, $\gamma^{\mu}_{\nu\beta} = \frac{g^{\mu\lambda}}{2}(\partial_{\nu}g_{\lambda\beta} + \partial_{\beta}g_{\lambda\nu} -\partial_{\lambda}g_{\nu\beta})$ is the Lorentzian Christoffel symbol \cite{chern} and $\tilde{\gamma}^{\mu}_{\nu\beta} = \frac{g^{\mu\lambda}}{2\sigma}(\partial_{\nu}s_{\lambda\beta} + \partial_{\beta}s_{\lambda\nu} -\partial_{\lambda}s_{\nu\beta})$. The equation (\ref{Bipartitesecondlaw}) is the generalized Newton's second law of motion with an anisotropic inertia and force, similar to found in \cite{Bogoslovsky:1998wa}.

For the flat Minkowsky space-time we find a new anisotropic 4-force
\begin{equation}
\tilde{F}^{\mu}=m\xi[(\tilde{\gamma}^{\mu}_{\nu\beta} + \partial_{\nu}(\sigma^{-1}))\delta^{\mu}_{\beta}]\dot{x}^{\nu}\dot{x}^{\beta}.
\end{equation}
Choosing the biparite tensor $\mathbf{s}=N^{2}(x)\mathbf{\eta}$ it turns out that the 4-force is given by
\begin{equation}
\tilde{F}^{\mu}=m\xi\Big[\partial^{\mu}N + \left(2-\frac{1}{N^{2}}\right)(\dot{x}^{\nu}\partial_{\nu}N)\dot{x}^{\mu}\Big]
\end{equation}
whose 3-force has the form
\begin{equation}
\label{3force}
\vec{\tilde{F}}=m\xi[\nabla N + \left(2-\frac{1}{N^{2}}\right)\left(\frac{\partial N}{\partial t} + (\vec{v}\cdot \nabla) N\right)\vec{v}].
\end{equation}
The choice for the particular form of the tensor $\mathbf{s}$ is inspired in the optical analogy between the light propagation in a medium with refraction index $N$ and in a curved space-time with a conformal metric $g_{\mu\nu}=N^{2}\eta_{\mu\nu}$ \cite{joetsis}. Then, we argue that for a static anisotropy and disregarding the quadratic term in the velocity, the anisotropy of Bipartite space induces an analog refraction index $N=\sigma$.
On the other hand, in this regime, the function $N$ can also be interpreted as an anisotropic potential. Thus, an intriguing new result is that a particle will suffer a deflection analogous to a charged particle in a static electric field.

The presence of the hydrodynamical derivatives $\frac{\partial}{\partial t}+\vec{v}\cdot \nabla$ suggests a mechanical analogy. We interpret the Bipartite force (\ref{3force}) as resulting from the interaction of the particle with a background fluid (aether) that changes its trajectory.

\section{Dynamics of the geometry}
\label{Dynamics of the geometry}
The analysis of the Bipartite geometry made so far was restricted to a fixed background situation. In this section we show that a dynamics for the Finslerian metric $\mathbf{g}^{F}$ provides, at least for the weak directional-dependence, a dynamics and interaction for the Lorentzian metric $\mathbf{g}$ and for the Bipartite tensor $\mathbf{s}$.

Consider the Finslerian Christoffel symbol $\gamma^{F\mu}_{\nu\beta} = \frac{g^{F\mu\lambda}}{2}(\partial_{\nu}g^{F}_{\lambda\beta} + \partial_{\beta}g^{F}_{\lambda\nu} -\partial_{\lambda}g^{F}_{\nu\beta})$ which can be written as $\gamma^{F\beta}_{\mu\nu} = \tilde{\gamma}^{F\beta}_{\mu\nu} + \hat{\gamma}^{F\beta}_{\mu\nu} + \bar{\gamma}^{F\beta}_{\mu\nu}$ where,
\begin{eqnarray}
\label{tildechristoffel}
\tilde{\gamma}^{F\beta}_{\mu\nu} & = & \gamma^{\beta}_{\mu\nu} + \frac{\alpha}{2F}\Big[\partial_{\mu}\left(\frac{F}{\alpha}\right)\delta^{\rho}_{\nu} + \partial_{\nu}\left(\frac{F}{\alpha}\right)\delta^{\rho}_{\mu} - \partial^{\rho}\left(\frac{F}{\alpha}\right)g_{\mu\nu}\Big]- \frac{\xi\alpha}{S}s^{\rho\epsilon}\gamma_{\epsilon\mu\nu}\nonumber\\
                      & - &  \frac{\xi\alpha^{2}}{2FS}\Big[\partial_{\mu}\left(\frac{F}{\alpha}\right)s^{\rho}_{\nu} + \partial_{\nu}\left(\frac{F}{\alpha}\right)s^{\rho}_{\mu} - s^{\rho\epsilon}\partial_{\epsilon}\left(\frac{F}{\alpha}\right)s_{\mu\nu}\Big],
\end{eqnarray}
with $\gamma^{\rho}_{\mu\nu}$ being the Lorentzian Christoffel symbols constructed from the Lorentzian metric $\mathbf{g}$,
\begin{eqnarray}
\label{hatchristoffel}
\hat{\gamma}^{F\beta}_{\mu\nu} & = & \frac{\xi\alpha}{2\sigma}[\nabla_{\mu}s^{\rho}_{\nu} + \nabla_{\nu}s^{\rho}_{\mu} - \nabla^{\rho}s_{\mu\nu} - 2s^{\rho}_{\epsilon}\gamma^{\epsilon}_{\mu\nu}]\nonumber\\
& - & \frac{\xi\alpha}{2F}\Big[\partial_{\mu}\left(\frac{F}{\sigma}\right)s^{\rho}_{\nu} + \partial_{\nu}\left(\frac{F}{\sigma}\right)s^{\rho}_{\mu} - \partial^{\rho}\left(\frac{F}{\sigma}\right)s_{\mu\nu}\Big] + \mathcal{O}(\xi^{2})
\end{eqnarray}
where $\nabla_{\mu}$ stands for the Levi-Civita connection compatible with the Lorentzian metric $\mathbf{g}$, and $\bar{\gamma}^{F\rho}_{\mu\nu}$ is formed by the derivatives of the vectors $k_{\mu}$.

Note that $\hat{\gamma}^{\rho}_{\mu\nu}$ has second order terms in $\xi$ whereas the $\bar{\gamma}^{\rho}_{\mu\nu}$ is explicit directional dependent, i.e., the directional dependence is present not only on the tensor components. Hereupon we shall take into account only the terms linear in $\xi$ and without explicit directional dependence. This regime shall be called a weak directional-dependence. The directional-dependence encoded only on the components of the tensor fields have already been addressed by other authors \cite{Pfeifer:2011xi,Pfeifer:2011tk}.

Let us choose the $\mathbf{g}^{F}$-compatible Cartan connection $\omega$ for the $TTM=\{(x,y),x\in M, y\in T_{x}M\}$. In order to do it we need to make the split $TTM= hTTM \oplus vTTM$, where $hTTM$ is the submanifold for $y$ fixed, called horizontal fiber whereas $vTTM$ is a fiber for $x$ fixed, called vertical fiber \cite{chern}. An orthornormal basis for $TTM$ is given by $(\frac{\delta }{\delta x^{\nu}},F\frac{\partial}{\partial y^{\nu}})$, where $\frac{\delta }{\delta x^{\nu}}=\frac{\partial}{\partial x^{\mu}}-N_{\nu}^{\beta}\frac{\partial}{\partial y^{\beta}}$ is a basis for $hTTM$ and $F\frac{\partial}{\partial y^{\nu}}$ is a basis for $vTTM$. The symbol $N^{\beta}_{\nu}$ is the so-called Nonlinear connection and is given by $ N_{\rho}^{\mu}= \gamma_{\mu\nu}^{\rho}y^{\nu} - \frac{A^{\rho}_{\mu\nu}}{F}\gamma^{\nu}_{\epsilon\xi}y^{\epsilon}y^{\xi}$. Taking the dual basis $(dx^{\nu},\delta y^{\nu})$ , where $\delta y^{\nu}=dy^{\nu} + N^{\nu}_{\beta}dx^{\beta}$, the connection takes the form $ \omega^{\rho}_{\mu} = \Gamma^{\rho}_{\mu\nu}dx^{\nu} + \frac{A_{\mu\nu}^{\rho}}{F}\delta y^{\nu}$, where $A_{\mu\nu\rho}= \frac{F}{2}\frac{\partial g^{F}_{\mu\nu}}{\partial y^{\rho}}$ is the so-called Cartan tensor and $ \Gamma^{\rho}_{\mu\nu} = \gamma^{F\rho}_{\mu\nu} - \frac{g^{F\rho\epsilon}}{F}(A_{\nu\epsilon\xi}N^{\xi}_{\\mu}-A_{\mu\nu\xi}N^{\xi}_{\epsilon})$ is the horizontal component \cite{chern,Pfeifer:2011tk,Vacaru:2010fa}. Since we restrict our analysis to the weak directional-dependence limit, we neglect the effects of the Cartan tensor what yields to
\begin{eqnarray}
\label{weakdependenceconnection}
\Gamma^{F\rho}_{\mu\nu} & = & \gamma^{\rho}_{\mu\nu}(x) + \frac{\xi\alpha}{2F}\Big[\partial_{\mu}\left(\frac{\sigma}{\alpha}\right)\delta^{\rho}_{\nu} + \partial_{\nu}\left(\frac{\sigma}{\alpha}\right)\delta^{\rho}_{\mu} - \partial^{\rho}\left(\frac{\sigma}{\alpha}\right)g_{\mu\nu}\Big]- \frac{\xi\alpha}{S}s^{\rho\epsilon}\gamma_{\epsilon\mu\nu}\nonumber\\
                      & - &  \frac{\xi\alpha^{2}}{2FS}\Big[\partial_{\mu}\left(\frac{F}{\alpha}\right)s^{\rho}_{\nu} + \partial_{\nu}\left(\frac{F}{\alpha}\right)s^{\rho}_{\mu} - s^{\rho\epsilon}\partial_{\epsilon}\left(\frac{F}{\alpha}\right)s_{\mu\nu}\Big]\nonumber\\
                      & + & \frac{\xi\alpha}{2\sigma}[\nabla_{\mu}s^{\rho}_{\nu} + \nabla_{\nu}s^{\rho}_{\mu} - \nabla^{\rho}s_{\mu\nu} - 2s^{\rho}_{\epsilon}\gamma^{\epsilon}_{\mu\nu}]\nonumber\\
& - & \frac{\xi\alpha}{2F}\Big[\partial_{\mu}\left(\frac{F}{\sigma}\right)s^{\rho}_{\nu} + \partial_{\nu}\left(\frac{F}{\sigma}\right)s^{\rho}_{\mu} - \partial^{\rho}\left(\frac{F}{\sigma}\right)s_{\mu\nu}\Big].
\end{eqnarray}
The Curvature 2-form is defined by $ R^{\delta}_{\alpha} = d\omega^{\delta}_{\alpha} + \omega^{\delta}_{\epsilon}\wedge\omega^{\epsilon}_{\alpha} = R_{\alpha\beta\gamma}^{F\delta}dx^{\beta}\wedge dx^{\gamma} + P_{\alpha\beta\gamma}^{\delta}dx^{\beta}\wedge \delta y^{\gamma} + Q_{\alpha\beta\gamma}^{\delta}\delta y^{\beta}\wedge \delta y^{\gamma}$ \cite{chern}.
In the weak directional-dependence limit we restrict ourselves to the horizontal-horizontal component of the curvature 2-form $R_{\alpha\beta\gamma}^{\delta}$
which has the familiar form $ R_{\alpha\beta\gamma}^{F\delta} = \delta_{\gamma}\Gamma^{\delta}_{\alpha\beta} - \delta_{\beta}\Gamma^{\delta}_{\alpha\gamma} + \Gamma^{\delta}_{\epsilon\beta}\Gamma^{\epsilon}_{\alpha\gamma} - \Gamma^{\delta}_{\epsilon\gamma}\Gamma^{\epsilon}_{\alpha\beta}$. Dropping the quadratic terms yields the Ricci tensor
\begin{eqnarray}
\label{Finslerianriccitensor}
R^{F}_{\mu\nu}  &  =  &   R_{\mu\nu} + \xi\Big\{\frac{\alpha}{2\sigma}(- \nabla_{\rho}\nabla^{\rho}s_{\mu\nu} + \nabla_{\mu\rho}s^{\rho}_{\nu} + \nabla_{\nu\rho}s^{\rho}_{\mu})\nonumber\\
                &  -  &  3\partial_{\mu}\left(\frac{F}{\alpha}\right)\partial_{\nu}\left(\frac{\alpha}{2F}\right) + \partial_{\nu}\left(\frac{F}{\alpha}\right)\partial_{\mu}\left(\frac{\alpha}{2F}\right) - 3\partial_{\mu\nu}\left(\frac{F}{\alpha}\right)\left(\frac{\alpha}{2F}\right)\nonumber\\
                &  -  &  g_{\mu\nu}\Big[\frac{\alpha}{2F}\Box\left(\frac{F}{\alpha}\right) + \partial_{\rho}\left(\frac{\alpha}{2F}\right)\partial^{\rho}\left(\frac{F}{\alpha}\right)\Big] - \frac{\alpha}{2F}\partial^{\rho}\left(\frac{F}{\alpha}\right)\partial_{\rho}g_{\mu\nu}\nonumber\\
                &  + & \partial_{\rho}\Big[\frac{\alpha^{2}}{2FS}[\partial_{\mu}\left(\frac{F}{\alpha}\right)s^{\rho}_{\nu} + \partial_{\nu}\left(\frac{F}{\alpha}\right)s^{\rho}_{\mu} - s^{\rho\epsilon}\partial_{\epsilon}\left(\frac{F}{\alpha}\right)s_{\mu\nu}]\Big]\nonumber\\
                &  -  &  \partial_{\nu}\Big[\frac{\alpha^{2}}{2FS}\left(s\partial_{\mu}\left(\frac{F}{\alpha}\right) + s^{\rho}_{\mu}\partial_{\rho}\left(\frac{F}{\alpha}\right) - \zeta s^{\epsilon}_{\mu}\partial_{\epsilon}\left(\frac{F}{\alpha}\right)\right)\Big]\nonumber\\
                &  - & \partial_{\rho}\Big[\frac{\alpha}{2F}\Big[\partial_{\mu}\left(\frac{F}{\sigma}\right)s^{\rho}_{\nu} + \partial_{\nu}\left(\frac{F}{\sigma}\right)s^{\rho}_{\mu} - \partial^{\rho}\left(\frac{F}{\sigma}\right)s_{\mu\nu}\Big]\Big]\nonumber\\
                &  + & \partial_{\nu}\Big[\frac{\alpha}{2F}\Big[\partial_{\mu}\left(\frac{F}{\sigma}\right)s + \partial_{\rho}\left(\frac{F}{\sigma}\right)s^{\rho}_{\mu} - \partial^{\rho}\left(\frac{F}{\sigma}\right)s_{\rho\mu}\Big]\Big]\nonumber\\
                &  -  & \partial_{\rho}\Big[\frac{\alpha}{S}s^{\rho\epsilon}\gamma_{\epsilon\mu\nu}\Big] + \partial_{\nu}\Big[\frac{\alpha}{S}s^{\rho\epsilon}\gamma_{\epsilon\mu\rho}\Big]  - 2\nabla_{\rho}(s^{\rho}_{\epsilon}\gamma^{\epsilon}_{\mu\nu})\Big\}.
\end{eqnarray}
The first line of eq.(\ref{Finslerianriccitensor}) is composed by second derivatives of both the Lorentzian metric $\mathbf{g}$ and of the Bipartite tensor $\mathbf{s}$ whereas the remaining lines have coupled terms. Thus, the first line provides propagators for the tensorial fields $\mathbf{g}$ and $\mathbf{s}$. Assuming the Finslerian Einstein equation $R^{F}_{\mu\nu}=\kappa\left(T^{F}_{\mu\nu}-\frac{T^{F}}{2}g^{F}_{\mu\nu}\right)$ holds, the Finlerian vacuum $R^{F}_{\mu\nu}=0$ equation can be interpreted as providing the dynamical equations for the tensorial field $\mathbf{s}$ by
\begin{equation}
\label{Bipartiteeom}
- \nabla_{\rho}\nabla^{\rho}s_{\mu\nu} + \nabla_{\mu\rho}s^{\rho}_{\nu} + \nabla_{\nu\rho}s^{\rho}_{\mu}=0
\end{equation}
and the Lorentzian Einstein equation $R_{\mu\nu}=\kappa\left(T_{\mu\nu}-\frac{T}{2}g_{\mu\nu}\right)$
whose source is given by the remaining coupled terms of the eq.(\ref{Finslerianriccitensor}).

The Equation of motion of $\mathbf{s}$ resembles the perturbed graviton equation which posses the gauge symmetry $s'_{\mu\nu}=s_{\mu\nu}+\nabla_{(\mu}\lambda_{\nu)}$. This is an important new result since the Randers vector also has a gauge symmetry $a_{\mu}'\sim a_{\mu}+\partial_{\mu}\Phi$ \cite{randers}. Choosing a Lorentz-like gauge $\nabla_{\mu}s^{\mu\nu}=0$, the dynamics of the Bipartite tensor $\mathbf{s}$ comes from the Lagrangian $\mathcal{L}_{s}=-\frac{1}{2}\nabla_{\rho}s^{\mu\nu}\nabla^{\rho}s_{\mu\nu}$.

The Bipartite geometry also yields the interaction terms between the Lorentzian metric $g_{\mu\nu}$ and the Bipartite tensor $s_{\mu\nu}$. Indeed, the Ricci scalar is given by
\begin{eqnarray}
R^{F}   &  =  &  g^{F\mu\nu}R^{F}_{\mu\nu}\nonumber\\
        &  =  &  \frac{\alpha}{F}R - \xi \frac{\alpha^{2}}{FS}s^{\mu\nu}R_{\mu\nu} + ... .
\end{eqnarray}
Further, the Jacobian determinants are related by \cite{Kostelecky:2012ac}
\begin{eqnarray}
\sqrt{|g^{F}|}  &   =   &   \left(\frac{F}{\alpha}\right)^{\frac{5}{2}}\left(\frac{S}{\sigma}\right)^{\frac{m-1}{2}}\sqrt{|g|}\nonumber\\
                &   =   &   \Big\{ 1 + \xi\Big[\frac{5}{2}\frac{\sigma}{\alpha} + \frac{(m-1)\zeta}{2}\frac{\alpha}{\sigma}\Big] + ...\Big\}\sqrt{|g|},
\end{eqnarray}
where $m$ is the multiplicity of the eigenvalue $\zeta$. Therefore, the Einstein-Hilbert Lagrangian yields
\begin{eqnarray}
\label{Bipartiteeinsteinhilbertaction}
\mathcal{L}_{EH}                         &   =   &   R^{F}\sqrt{|g^{F}|}\nonumber\\
                                    &   =   &   \Big\{R + \Big[\frac{3}{2}\frac{\sigma}{\alpha} + \frac{(m-1)}{2}\frac{\alpha}{\sigma}\Big]\xi R - \xi\frac{\sigma}{\alpha}s^{\mu\nu}R_{\mu\nu} + ... \Big\}\sqrt{|g|}.
\end{eqnarray}
which is analogous to the interaction terms of the gravitational sector of the Standard Model Extension \cite{Kostelecky:2003fs,Maluf:2013nva}. Hence, a Finslerian geometric dynamics in the weak directional dependence limit can be view as a Lorentzian geometry interacting with a Lorentz-Violating background tensor field $\mathbf{s}$.


\section{Final remarks and perspectives}
\label{Final remarks and perspectives}

In this letter we found new and interesting features of the Bipartite space which we outline some additional comments.

As shown in eq.(\ref{generalcausalsurface}), we find that the Bipartite tensor $\mathbf{s}$ deforms the causal cone stretching or squeezing it according to the sign of the eigenvalues of $\mathbf{s}$. Moreover, the deformation is of second order in $\xi$ and the condition (\ref{causalcondition}) on $\mathbf{s}$ guarantees that the perturbed cone lies inside the unperturbed one. However, other Finsler structures, as studied in great details in \cite{Pfeifer:2011xi,Pfeifer:2011tk}, reveal faster than light speeds as natural consequences. These results do not contradict themselves since the Finsler structure proposed in these works are different. Furthermore, the Randers space possesses a double cone as causal surfaces \cite{Chang:2008zzb}. Therefore, we conclude that the causal structure of the space-time is rather sensible to the Finsler geometry chosen.

From Eq.(\ref{bipartitegamma}) we conclude that that the Lorentz transformations are altered by the Bipartite tensor. Since the geometry is defined on $TTM$ a straight perspective is find the generalized transformation of the coordinates $(x^{\mu},y^{\mu})$ of $TTM$ that keep the Bipartite structure invariant. This approach reveals that a Lorentz Violation on $TM$ can be considered as a result of a bigger symmetry on $TTM$.

The 4-momentum we find in eq.(\ref{finslermomentum}) is not parallel to the 4-velocity.
This result is analogous to anisotropic crystals
where the displacement and the electric vectors are related by $D^{i}=\epsilon^{i}_{j}E^{j}$, where $\epsilon_{ij}$ is the permissivity tensor. This gives rise to the birefringence phenomenon
which is predicted by Lorentz-Violating theories in Minkowsky space-time \cite{Colladay:1998fq} and due to the anisotropic effects in Finsler space-times \cite{Skakala:2008jp}. Thus, we argue that the Bipartite tensor $\mathbf{s}$
can be interpreted as an analogous dieletric tensor of the anisotropic space-times. A better description of these electromagnetic phenomena through the coupling of the vector gauge field and the anisotropic Finsler metric, as done by Pfeifer and Wohlfarth in \cite{Pfeifer:2011tk}, is left as a perspective.

The coupling of the particle with the Finsler metric we find in eq.(\ref{modifieddispersionsrelations}) yields to modification of the dispersion relations analogous to the non-standard kinetic terms important to cosmology \cite{RenauxPetel:2008gi} and to topological defects \cite{Bazeia:2012rc}. Furthermore, the quartic terms in the momentum can yield ELKO spinors whose dispersion relation is quadratic \cite{Ahluwalia:2004ab,Bernardini:2012sc}. This exotic spinor is a candidate for the dark matter \cite{Ahluwalia:2004ab}. We postpone to a future work a complete analysis of dynamics of fields on a Finslerian space-time which has to be define on $TTM$ and so take into account the directional derivatives, as done by Pfeifer and Wohlfarth \cite{Pfeifer:2011tk}.

The anisotropic force in eq.(\ref{Bipartitesecondlaw}) obtained here can also leads to intriguing new features as the light bending around a massive star or even an analogous black holes where the light would be trapped due to the anisotropy. In order to study these conjectured effects we left as a next step the analysis of the gravitational equations by means of the osculating Riemannian space approach of Kouretsis \textit{et al} \cite{Kouretsis:2008ha,Kouretsis:2010vs} or using the method developed by Chang and Xi \cite{Chang:2008yv,Chang:2009pa}.

In the analysis of the Finslerian Einstein equations we show that the geometry has a dynamics similar to a Lorentzian one with a background dynamical tensor field $\mathbf{s}$ as a source. It is a perspective to seek for a general gauge symmetry for the Finslerian metric $\mathbf{g}^{F}$, through the Killing vector, which induces a gauge invariance for both $\mathbf{g}$ and $\mathbf{s}$ for tiny $\xi$. Further, we intend to go beyond the weak directional-dependence limit by studding the whole Einstein equation on $TTM$. Moreover, we would like to study the effects provided by the vertical-vertical $P_{\alpha\beta\gamma}^{\delta}$ and horizontal-vertical $Q_{\alpha\beta\gamma}^{\delta}$ curvature components.

Another improvement of the present work refers to phenomenological consequences of this model which leads to bounds on the Bipartite parameter $\xi$. In this regard we argue that the best samples from the particle Physics, as the decay of particles, needs the description of the deformed Lorentz-Bipartite transformations and the coupling between fields and Finsler metric, as discussed above. From gravitation and cosmology, besides the deformed Lorentz transformations, a deeper analysis of the Finslerian Einstein equations is required to obtain the light bending, for instance. All of these important effects are in order to augment the Bipartite model but due to their complexity they should be treated in a future work.


\section{Acknowledgments}

The authors are grateful to Alan Kosteleck\'{y} and Yuri Bonder for useful discussions and to Brazilian agencies CNPq and FUNCAP for financial support.


\end{document}